\pdfoutput=1
\documentclass[aps,prb,reprint,twocolumn,footinbib]{revtex4-2}

\usepackage{graphicx}
\usepackage{bm}
\usepackage{hyperref}
\usepackage{amsmath}
\usepackage{amssymb}
\usepackage{mathtools}
\usepackage{siunitx}
\usepackage{multirow}

\makeatletter
\g@addto@macro\bfseries{\boldmath}
\makeatother

\usepackage{comment}
\usepackage[dvipsnames,svgnames]{xcolor}
\usepackage{array}
\usepackage{multirow}
\usepackage{physics}
\usepackage[compat=1.0.0]{tikz-feynman}
\usepackage[caption=false]{subfig}
\usepackage{braket}

\usepackage[english]{babel}
\usepackage[autostyle, english = american]{csquotes}

\usepackage[scr=kp]{mathalpha}

\makeatletter
\def\maketitle{
\@author@finish
\title@column\titleblock@produce
\suppressfloats[t]}
\makeatother

\newcommand{\beq}{\begin{equation}}
\newcommand{\eeq}{\end{equation}}
\newcommand{\subhead}[1]{\textit{#1.---}}

\newcommand{\J}{C}

\newcommand{\bs}{\boldsymbol}

\begin{document}

\title{{Charge-$2e$ superconductivity from a disordered pair density wave}}

\author{Julian May-Mann}
\email{maymann@stanford.edu}
\affiliation{Department of Physics, Stanford University, Stanford, California 94305, USA}

\author{Akshat Pandey}
\email{akshatp@stanford.edu}
\affiliation{Department of Physics, Stanford University, Stanford, California 94305, USA}

\author{Zhengyan Darius Shi}
\email{zhengyanshi@stanford.edu}
\affiliation{Department of Physics, Stanford University, Stanford, California 94305, USA}

\author{Steven A.~Kivelson}
\email{kivelson@stanford.edu}
\affiliation{Department of Physics, Stanford University, Stanford, California 94305, USA}

\date{\today}

\begin{abstract}

We investigate the effects of disorder on a system that in the clean limit is a pair density wave (PDW) superconductor. The charge order of the clean PDW is inevitably lost (via Imry-Ma), but the fate of the superconducting order is less clear. Here, we consider a strongly inhomogeneous limit in which the system consists of a random collection of PDW puddles embedded in a metallic background. When the puddles are dilute, they become phase coherent at low temperatures, resulting in a state that is macroscopically equivalent to a charge-$2e$ $s$-wave superconductor. This is a mechanism by which a PDW can have zero resistance, even in the presence of disorder, and is thermodynamically distinct from the ``vestigial'' charge-$4e$ superconductivity that has been proposed to arise in weakly disordered PDWs.

\end{abstract}

\maketitle

\subhead{Introduction}
Quenched randomness is generally detrimental to the existence and strength of long-range order. The Imry-Ma argument implies that arbitrarily weak random fields prevent the formation of long-range order below a critical
dimension of $d = 2$ for discrete and $d = 4$ for continuous symmetries~\cite{ImryMa1975, AizenmanWehr, larkin1970effect}. However, the effects of disorder can be subtle when multiple ordering tendencies are intertwined~\cite{fradkin2015colloquium}, potentially leading to qualitatively new ordered phases that are not realized in the clean system~\cite{cho2018using, leroux2019disorder, NieTarjusKivelson, cui2018smeared, fernandes2019intertwined}.

A natural problem to consider in this regard is the fate of pair density wave (PDW) superconductors in the presence of randomly located charged impurities. PDWs are an exotic form of superconductivity where the phase and/or amplitude of the superconducting gap periodically oscillates in space, while its spatial average vanishes~\cite{PDWreview}.  Although PDWs were initially considered in the context of cuprate superconductors~\cite{berg2007dynamical,wang2018pair, du2020imaging}, they have since been argued for in Kagome~\cite{chen2021roton,deng2024chiral}, heavy fermion~\cite{aishwarya2023magnetic, gu2023detection, aishwarya2024melting}, iron-based~\cite{liu2023pair}, and multi-layer graphene~\cite{han2025signatures} systems.
A number of toy theoretical models also show PDW ground states~\cite{berg2010pair, jaefari2012pair, may2020topology, wu2023pair, yerin2023multiple, liu2024pair}.

Here, we focus on unidirectional Larkin-Ovchinnikov type PDWs~\cite{larkin1965nonuniform}, for which the gap function can be written as
\begin{equation}
    \Delta(\bm{r}) = \Delta_{\bm{Q}}e^{i \bm{Q}\cdot \bm{r}} + \Delta_{-\bm{Q}}e^{-i \bm{Q}\cdot \bm{r}},
\label{eq:PDWDef}\end{equation}
where $\bm{r}$ is the center-of-mass coordinate of the Cooper pair and $\bm{Q}$ is the PDW ordering vector, which we take to be incommensurate with the lattice~\footnote{In $d=2$ a commensurate order parameter also suffices for Imry-Ma purposes.}. It is straightforward to generalize to PDWs with multiple ordering vectors.

\begin{figure}[t!]
\centering
    \includegraphics[width=.7\linewidth]{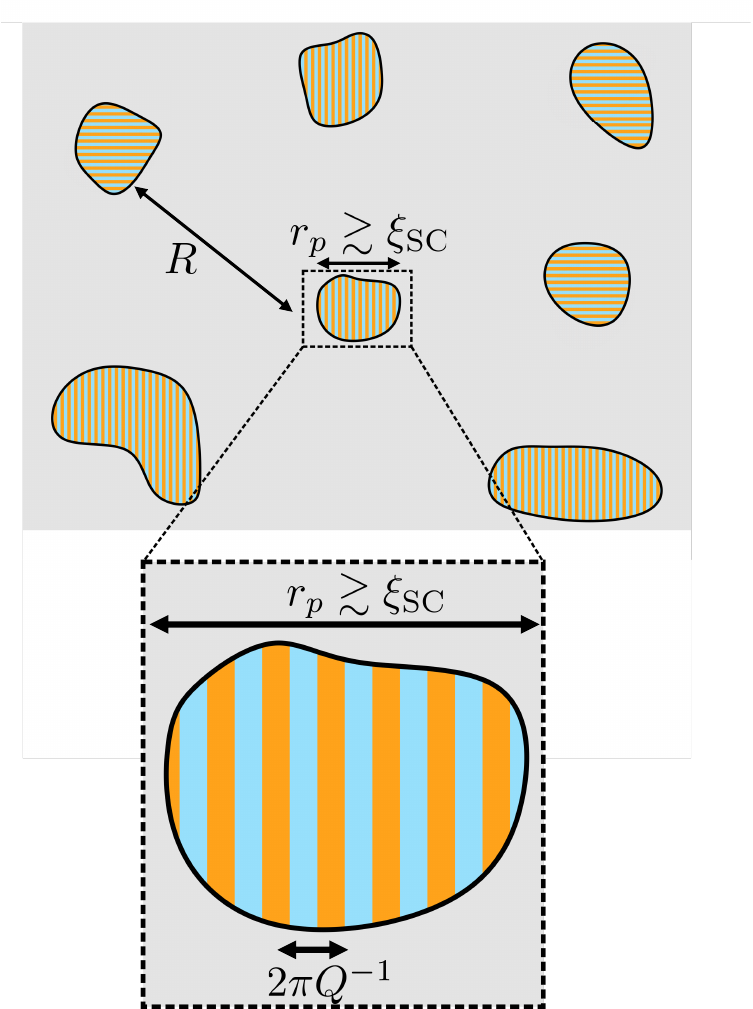}
    \caption{Relevant length scales for the PDW puddles: the interpuddle distance $R$, the characteristic puddle size $r_p$, the superconducting correlation length $\xi_{\text{SC}}$, and the PDW wavelength $Q^{-1}$.
    }
    \label{fig:scales}
\end{figure}

\begin{table}
\caption{Length scales considered in this Letter and their hierarchy. In this regime, the disordered PDW puddles realize a macroscopic $s$-wave superconducting phase.}
\label{tab:scales}
\begin{ruledtabular}
\begin{tabular}{lc}
\multicolumn{2}{c}{Length Scales} \\
\hline
PDW wavelength: & $Q^{-1}$\\
PDW correlation length: & $\xi_{\text{SC}}$\\
Characteristic PDW puddle size: & $r_p$\\
Inter PDW puddle distance: & $R$\\
Metal correlation length: & $L_T$\\
\hline
\multicolumn{2}{c}{Hierarchy of Scales} \\
\hline
\multicolumn{2}{c}{$Q^{-1} , \ \xi_{\rm SC} \lesssim r_p   \ll R \lesssim L_T$}\\
\end{tabular}
\end{ruledtabular}
\end{table}

Before discussing the effects of random impurities on PDWs, it will be useful to first discuss the effects of random impurities on charge density waves (CDWs) and uniform superconductivity. Random impurities, and related forms of disorder, induce random field couplings for CDWs, preventing long range order below the relevant critical dimension. By contrast, these forms of disorder do not lead to any random field couplings for uniform superconductivity as such couplings would break gauge invariance. Uniform superconductivity is therefore stable in the presence of disorder, while CDWs are not. Let us now turn our attention to PDWs. As with uniform superconductors, impurities do not induce random field couplings for PDWs. Nevertheless, they can still
affect PDWs through a random field type coupling to the associated composite CDW order parameter, $\rho_{2\bm Q} =  \Delta_{\bm Q} \Delta^*_{-\bm Q}$~\cite{agterberg2008dislocations, berg2009theory}. The Imry-Ma arguments tell us that $\rho_{2\bm Q}$ cannot order in the presence of disorder, and this necessarily implies that $\Delta_{\pm \bm Q}$ cannot individually order either. With this in mind, there have been two main proposals for the fate of PDWs in the presence of disorder~\cite{berg2007dynamical, MrossSenthil,chan2016interplay}:
either the superconducting  (SC) order is destroyed along with the composite CDW order or all charge-$2e$ SC orders are destroyed but a composite uniform charge-$4e$ SC order parameter $\Delta_{4e} = \Delta_{\bm Q} \Delta_{-\bm Q}$ persists as ``vestigial'' order.

Here, we show that there is a third possible fate, where a disordered PDW becomes \textit{macroscopically} equivalent to a conventional charge-$2e$ $s$-wave superconductor. One way to understand why this state is unexpected is to view the PDW order as composite order, which can be expressed as the product of two new emergent order parameters, \begin{equation}
    \Delta_{\pm \bm{Q}}({\bm r}) = \Delta_0({\bm r}) \rho_{ \pm\bm{Q}}({\bm r}) ,
\label{eq:fracPDW}\end{equation}
 where $\Delta_0$ is the order parameter of a uniform charge-$2e$ s-wave superconductor and $\rho_{\bm{Q}}\equiv \rho^\star_{-\bm{Q}}$ is a wave vector-$\bm{Q}$ CDW. In this rewriting, we have effectively separated the momentum and charge carried by the PDW order parameter~\cite{komijani2018order}. Individually, $\Delta_0$ and $\rho_{\bm{Q}}$ do not constitute local order parameters as there is a $\mathbb{Z}_2$ gauge symmetry associated with this decomposition; all physical observables are invariant under $\Delta_0(\bm{r}) \to s({\bm r})\Delta_0(\bm{r})$ and $\rho_{\bm{Q}}(\bm{r}) \to s({\bm r})\rho_{\bm{Q}}(\bm{r})$ where $s(\bm{r})=\pm 1$.  In the clean PDW phase, neither $\Delta_0$ nor $\rho_{\bm{Q}}$ orders---only the original PDW and composites thereof can order.

 Nevertheless, we show here that, with disorder, a PDW system can enter a superconducting phase that is equivalent to a charge-$2e$ s-wave superconductor in macroscopic phase sensitive measurement and, moreover, that this superconducting phase is consistent with the $\mathbb{Z}_2$ gauge structure discussed above.  This scenario, which we shall refer to as the ``macroscopic $s$-wave" scenario, is fundamentally distinct from the vestigial charge-$4e$ scenario considered in the existing literature.

In this Letter, we provide an explicit construction for realizing this macroscopic $s$-wave state in a system with quenched disorder.
Our starting point is a strongly inhomogeneous system, with the PDW fractured into a collection of far-separated puddles (equivalently, grains)~\cite{larkin1972density, ghosal2001inhomogeneous, dodaro2018generalization}. Each puddle hosts a local PDW order parameter which  couples to disorder via the local composite CDW order, $\rho_{2\bm{Q}}$.
When the disorder is strong, it pins the relative phases of $\Delta_{\bm{Q}}$ and $\Delta_{-\bm{Q}}$ (i.e., the CDW phase) to a random value at each puddle. However, the average of the two phases (the ``superconducting phase'') on each puddle is unpinned, as required by gauge invariance; the superconducting phases on neighboring puddles can therefore be Josephson coupled to one another via the metallic background. In the limit of dilute puddles, the dominant Josephson couplings are disordered but almost entirely unfrustrated, similar to the Mattis model~\cite{Mattis}. From this, we show that at low enough $T$, a charge-$2e$ globally $s$-wave SC phase arises. In terms of the $\mathbb{Z}_2$ gauge structure, this phase can be understood as arising from a disorder-induced Higgsing of the $\mathbb{Z}_2$ gauge field. Were this scenario to arise in a real material, it would behave like a conventional $s$-wave superconductor in all macroscopic phase sensitive measurements~\cite{van1995phase}. However, a local probe, like scanned Josephson tunneling microscopy, would only find PDW order~\cite{chen2022identification}.

An important consequence of this Letter applies to the transport signatures of pure PDWs (i.e., PDWs that do not coexist with a uniform superconductor). Previous theoretical analysis in Ref.~\cite{berg2007dynamical} concluded that the most natural expectation is that a pure PDW would not support a zero resistance state in a real material, where disorder is always present, unless a highly exotic charge $4e$ phase emerges. Our Letter demonstrates that there is a mechanism by which a pure PDW can display zero resistance while remaining a conventional $2e$ phase. This is important for interpreting transport signatures in any material that in the ideal (zero disorder) limit is conjectured to host a pure PDW phase. Additionally, it should be possible to induce a transition to the macroscopic $s$-wave phase found here by increasing disorder in a clean PDW system using, for example, chemical substitution.

A similar construction of a globally $s$-wave state for dilute grains of $d$-wave superconductor was discussed in Refs.~\cite{SpivakOretoKivelson, KivelsonSpivak}.
Following the logic used in the introduction, we can understand this state by treating the $d$-wave superconductor as the product of an $s$-wave superconductor, which is ordered, and a nematic, which is destroyed by disorder.

\subhead{Pair density wave puddles}
The starting point for our analysis is a collection of finite-sized puddles embedded in a weakly disordered Fermi liquid in $d = 2$ or $3$ dimensions. Puddles, as used in this Letter, correspond to well-localized mean-field solutions to the superconducting gap equation. This is natural to consider in situations where disorder leads to an emergent granularity. Additionally, localized solutions are generic for strongly disordered superconductors with short coherence lengths, where they appear as Lifshitz tails in the spectrum of the linearized gap equation~\cite{dodaro2018generalization}. There are then a number of relevant length scales: the PDW wavelength $Q^{-1}$, the clean-limit superconducting coherence length $\xi_{\rm SC}$, the average puddle size $r_p$,  the average interpuddle distance $R$, and the coherence length of the metal, $L_T = \min(v_F/T, \sqrt{D/T})$, where $v_F$ is the Fermi velocity and $D$ is the $T=0$ diffusivity due to impurities. These length scales are summarized in Table~\ref{tab:scales} and shown schematically in Fig.~\ref{fig:scales}.
Here, we adopt the following hierarchy of scales:
$Q^{-1} , \ \xi_{\rm SC} \lesssim r_p   \ll R \lesssim L_T$.
To have a well-defined CDW ordering vector, it is necessary that $Q^{-1} \lesssim r_p$, while $\xi_{\rm SC} \lesssim r_p$
is needed for the puddles to exist as the solution to some mean-field gap equation~\cite{muhlschlegel1972thermodynamic}.
The dilute limit of puddles is represented by $r_p\ll R$.  Finally, $R \lesssim L_T$  is necessary in order to have significant Josephson couplings between puddles and is always satisfied at low temperatures due to the $T \rightarrow 0$ divergence of $L_T$.  We also assume that the character of disorder is such that the probability of finding a dislocation of the composite CDW order within a puddle is small. However, including dislocations does not change any of our central results, as we shall discuss later.

Given these conditions, we can analyze the low temperature properties of the system by approximating the PDW phases and amplitude to be spatially constant within each puddle. The local gap function for the $i^{\text{th}}$ puddle is then
\begin{equation}\begin{split}
   \Delta_i(\bm{R}_i + \bm r) &=  f_i(\bm{r}) |\Delta_i| \left[ e^{i \theta_{i,+}} e^{i\bm{Q}_i \cdot \bm{r}} + e^{i \theta_{i,-}} e^{-i\bm{Q}_i \cdot \bm{r}} \right]\\
    &=  2f_i(\bm{r}) |\Delta_i|\cos(\bm{Q}_i\cdot \bm{r} + \phi_i)e^{i \theta_i}.
\label{eq:PDWPuddleOP}\end{split}\end{equation}
where we define the total and relative PDW phases as
\begin{equation}
    \theta_i = \frac{\theta_{i,+} + \theta_{i,-} }{2} \, \text{ and } \, \phi_i = \frac{\theta_{i,+}  - \theta_{i,-}}{2}
\label{eq:phaseReDef}\end{equation}
respectively. Here $\bm{R}_i$ is the center of puddle; $f_i$ encodes the size and shape of the puddle [$f_i({\bm 0}) = 1$ and $f_i({\bm r}) \to 0$ for $|{\bm r}| \gtrsim r_p$]; and
$|\Delta_i|$ and $\bm{Q}_i$ are the PDW amplitude and ordering vector, respectively.
We note in passing that $2\theta_i$ is the phase of the $4e$ SC composite order parameter and $2\phi_i$ is the phase of the wave vector $2\bm{Q}$ composite CDW, although these composite orders will not be relevant for the present analysis. Equation~\eqref{eq:phaseReDef} also induces a $\mathbb{Z}_2$ gauge redundancy as the local shift $(\theta_i,\phi_i) \rightarrow (\theta_i+\pi,\phi_i+\pi)$ leaves the original PDW phase variables, $\theta_{i,\pm}$, invariant. This is the same $\mathbb{Z}_2$ redundancy discussed in the introduction.

Because of disorder, $|\Delta_i|$ and $\bm{Q}_i$ are random, as are the locations of the puddles and hence the distances $|\bm{R}_i-\bm{R}_j|$ between neighboring puddles $i$ and $j$. The random puddle shapes defined by $f_i(\bm r)$ induce a distribution on their sizes $r_{p,i}$, defined by $r^d_{p,i} \sim \int d^d\bm{r}f_i(\bm{r})$.
We assume that none of the aforementioned scalar random variables have distributions with heavy tails.

\subhead{Coupling to disorder and Josephson couplings}
The partial ordering of the PDW puddles is governed by the pinning of the local composite CDW order by disorder on each puddle and by the Josephson couplings between pairs of puddles. The intrapuddle coupling to disorder can be expressed as
\begin{equation}\begin{split}
    H_{\text{dis}} &= -\sum_i h_i \Delta_{i,-\bm{Q}}\Delta^*_{i,+\bm{Q}} + c.c.\\
    &=-\sum_i 2|h_i||\Delta_i|^2\cos(2\phi_i - \alpha_i)
\end{split}\label{eq:disorderCoup}\end{equation}
where $h_i = |h_i|e^{i\alpha_i}$ is a random complex scalar that encodes the disorder potential at puddle $i$.
When disorder is strong (as we assume it is here), $\phi_i$ will be pinned to one of the two minima of the cosine potential, $\langle \phi_i \rangle = \alpha_i/2 + \pi (\tau_i - 1)/2$, where $0\leq \alpha_i/2 < \pi$ and $\tau_i = \pm 1$ is a remaining Ising degree of freedom at site $i$. Neither $\theta_i$ nor $\tau_i$ is pinned by disorder.

The Josephson couplings between different PDW puddles is mediated by the propagation of Cooper pairs through the weakly disordered Fermi liquid. The coupling between the PDW puddles and the Fermi liquid is given by
\begin{equation}
    H_{\text{tun}} =  \sum_i \!\! \int\!\! d^d \bm r   \Delta_i
    ( \bm{r}) \psi^\dagger_{\uparrow}(\bm{r})\psi^\dagger_{\downarrow}(\bm{r}) + h.c.,
\end{equation}
where $\psi_{\updownarrow}$ are the electron operators.
If $|\Delta_i|$ is small, the Josephson couplings can be computed by integrating out the Fermi liquid perturbatively. Upon doing this, we arrive at the following low-energy effective Hamiltonian for the remaining $\theta$ and $\tau$ degrees of freedom:
\begin{equation}
    H_{\rm eff}[\{ \theta_i\}, \{\tau_i\}] = - \sum_{ij} J_{ij} \tau_i \tau_j \cos(\theta_i - \theta_j),
\label{eq:IsingXYmodel}\end{equation}
Here, the $\mathbb{Z}_2$ gauge symmetry corresponds to the local transformation $(\theta_i,\tau_i) \rightarrow (\theta_i+\pi,-\tau_i)$. This Hamiltonian has a similar structure to that of the $\mathbb{Z}_2$ gauged XY-model with
$U_{ij}\equiv \tau_i \tau_j$ playing the role of the $\mathbb{Z}_2$ gauge connection~\cite{fradkin1979phase, lammert1993topology, sachdev2019topological}.
 However, this comparison should be used with caution. If $U_{ij}$ did actually reflect a dynamical gauge field, there would be configurations where the product of $U$'s around a given loop equals $-1$, ($\prod_{loop} U_{ij}=-1$). Here, it is straightforward to show that any such loop product would equal $+1$, and Eq.~\ref{eq:IsingXYmodel} is therefore distinct from the $\mathbb{Z}_2$ gauged XY-model.
 On the other hand, as we show in the Supplemental Material (SM) below, Eq.~\ref{eq:IsingXYmodel} is, indeed, related to a somewhat different gauge theory involving both XY and Ising degrees of freedom, both coupled to the same $\mathbb{Z}_2$ gauge connection. We note that Ref.~\cite{MrossSenthil} also considered this model, but with different disorder couplings, and found a $4e$ phase instead of a $2e$ phase. The similarities and differences between these constructions is discussed in the SM below.

The couplings in Eq.~\ref{eq:IsingXYmodel} can be calculated perturbatively in powers of $|\Delta|$. For $|\bm{R}_i - \bm{R}_j| \ll L_T$ the Josephson coupling, $ J_{ij}$ is
\begin{equation}\begin{split}
    J_{ij} = 4\nu &|\Delta_i||\Delta_j| \int d^d\bm{r}_i d^d\bm{r}_j f_i(\bm{r}_i)f_j(\bm{r}_j) \\&\times \frac{\cos(\bm{Q}_i\cdot \bm{r}_i + \alpha_i/2)\cos(\bm{Q}_j\cdot \bm{r}_j + \alpha_j/2)}{|\bm{R}_i + \bm{r}_i - \bm{R}_j - \bm{r}_j|^d},
\label{eq:JC_def}\end{split}\end{equation}
where $\nu$ is the density of states at the Fermi level~\footnote{Even in the presence of weak elastic scattering, so long as $k_F\ell\gg 1$, Eq. \ref{eq:JC_def} applies for the configuration averaged Josephson coupling. }.
The pinning fields, $\alpha_i$ and $\alpha_j$, enter the Hamiltonian due to the pinning of $\phi_i$ and $\phi_j$ by Eq.~\eqref{eq:disorderCoup}. For puddles where $|\bm{R}_i - \bm{R}_j|$ is greater than $L_T$, the Josephson couplings are exponentially suppressed.

To simplify the nontrivial Josephson couplings in Eq.~\ref{eq:JC_def}, we perform a multipole expansion:
\begin{equation}\label{eq:expansion}
\begin{split}
    J_{ij} =  4\nu |\Delta_i||\Delta_j| \sum^{\infty}_{\ell, \ell' = 0}  \frac{\J^{\ell,\ell'}_{ij}}{|\bm{R}_i - \bm{R}_j|^{d+\ell+\ell'}}
\end{split}\end{equation}
where $\J^{\ell,\ell'}_{ij}$ corresponds to the coupling of the $\ell^{\text{th}}$ multipole of the PDW on puddle $i$
to the $\ell'^{\text{th}}$ multipole of the PDW on puddle $j$.
This series is convergent in the dilute limit, $r_p \ll R$ and for sufficiently well-localized puddles.
To be more explicit, we define the rank-$\ell$ tensors $\eta^\ell_i$
\begin{equation}\begin{split} &\eta_{i,\left[\alpha,\beta,\ldots\right]}^\ell=\int d^d\bm{r} \ r_{\alpha}r_{\beta} \dots\ f_i(\bm{r}) \cos(\bm{Q}_{i}\cdot \bm{r} + \alpha_i/2),
\label{eq:etaGen}\end{split}\end{equation}
where $\alpha,\beta,\dots$ are the $\ell$ tensor indices, and $r_{\alpha}$ is the corresponding component of $\bm{r}$. The leading-order couplings, $\J^{0,0}$, $\J^{1,0}$ and $\J^{0,1}$, can then be written in terms of the ``charge'' $\eta^{0}$ and ``dipole moment'' $\bm{\eta}^{1}$ as
\begin{equation}
\begin{gathered}
\J^{0,0}_{ij} = \eta^0_i \ \eta^0_j,
\\ \J^{1,0}_{ij}
=\J^{0,1}_{ji}=-\frac{d}{2}\left[
{\bm\eta}^1_i\cdot{\bm n}_{ij}\right]\ \eta^0_j
,
\end{gathered}
\label{eq:coeffDef}\end{equation}
where $\bm{n}_{ji} = -\bm{n}_{ij} = \frac{\bm{R}_i-\bm{R}_j}{|\bm{R}_i-\bm{R}_j|}$ is the unit vector connecting puddles $i$ and $j$. Importantly, $ \J^{\ell,\ell'}_{ij}$ is independent of the distance between the puddles, $|{\bm R}_i- {\bm R}_j|$.

In the dilute limit $r_p/R  \ll 1$, the leading order coupling is the $(\ell,\ell') = (0,0)$ coupling. This coupling has a random amplitude and sign due to the random value of $\alpha_i/2$ in Eq.~\eqref{eq:etaGen}. However, despite the random signs, the $(0,0)$ couplings are secretly \textit{unfrustrated}. This can be made apparent by redefining the $\tau$ and $\theta$ variables as
\begin{equation}\begin{split}\label{eq:gauge}
    &\tau_i \rightarrow \tau'_i = \tau_i \ \text{sgn}[ \eta^0_i],\\
    &\theta_i \rightarrow \theta'_i = \theta_i + \pi (\tau'_i - 1)/2.
\end{split}\end{equation}
This amounts to a convenient choice of $\mathbb{Z}_2$ gauge, and makes all the $(0,0)$ couplings positive. (The amplitudes remain random.)

To the extent that higher order, $(\ell, \ell') \neq (0,0)$, couplings can be ignored, the present problem is equivalent to an XY ferromagnet that acts on local variables $|\Delta_i| e^{i \theta'_i} = |\Delta_i| \text{sgn}[\eta^0_i]\tau_i e^{i \theta_i}$.
Necessarily, in $d \geq 2$, it undergoes a transition to an ordered ($d>2$) or quasi-long-range ordered ($d=2$) phase below a nonzero $T_c$~\footnote{There are subtleties associated with the fact that $\int d^d R/R^d$  diverges logarithmically, so the way the Josephson coupling is cut off at long distances is always significant.  At finite $T$, the relevant length scale is $L_T$ so that, up to logarithmic corrections,  $T_c\sim (r_p/R)^d \nu | \Delta|^2$, where $| \Delta|$ is the typical magnitude of the gap function on a puddle.  Far below $T_c$, as $L_T \to \infty$, the perturbative computation  of $J_{ij}$ breaks down when a proximity effect induced gap in the metallic matrix serves to cut off the Josephson couplings at long distances.} Since $\tau_i e^{i \theta_i}$ preserves the $\mathbb{Z}_2$ gauge symmetry but breaks charge conservation symmetry, we conclude that the ordered phases is a superconductor.

While we have only considered pair-density wave puddles of the form given in Eq.~\eqref{eq:PDWPuddleOP}, the multipole expansion used here can be applied to \textit{any} type of superconducting puddle, provided that the coupling between the puddles and the normal Fermi liquid can be treated perturbatively. One relevant situation to consider is when the stripe order of individual puddles is distorted by dislocations of the composite CDW order. Nearly identical analysis shows that, even with distortion, the $(0,0)$ couplings are also dominant in this regime, and all other conclusions follow \textit{mutatis mutandis}.

\subhead{Phase sensitivity}
We now address the behavior of the superconducting state, as
would be detected in phase-sensitive measurements~\cite{geshkenbein1987vortices, van1995phase}. Consider an experiment on a device consisting of a sample of an SC of unknown symmetry connected at two edges to macroscopic leads of
$s$-wave SCs: SC 1 and SC 2, with phases $\Theta_1$ and $\Theta_2$ respectively. The macroscopic character of the SC state of the sample---e.g., $s$-wave vs $d$-wave or PDW vs uniform---can be determined from the $\Theta_1-\Theta_2$ dependence of the total supercurrent through the device for different sample device sizes and geometries.

The coupling between puddles and the SC leads is governed by the same physical processes as the coupling between puddles. Here, it is useful to separate the couplings between the leads and the puddles into two groups: the couplings to distant puddles that are $\gg r_p$ away from the leads and the couplings to close-by puddles. In the SM below we show that, for distant puddles, the Josephson couplings are again determined by the charge moment of the puddles, $\eta^0$, and are rendered ferromagnetic by Eq.~\ref{eq:gauge}. However, in this gauge, the couplings to the close-by puddles have random signs and are expected to be large. Nevertheless, the couplings to the nearby puddles can be neglected for macroscopic leads as the total energy associated with the nearby couplings is $\propto \sqrt{A_{\text{lead}}}$ where $A_{\text{lead}}$ is the cross-sectional area of the lead, while the total energy associated with the couplings to the distant puddles is $\propto A_{\text{lead}}$. The PDW puddles are thus equivalent to a set of $2e$ $s$-wave grains---supercurrent is proportional to $\sin(\Theta_1-\Theta)$ with a magnitude (critical current) that scales with the area of the junctions to the leads.

\subhead{Finite puddle densities and frustrated rare regions}
The asymptotic simplicity that arises from neglecting all  higher multipoles in Eq.~\ref{eq:expansion} is related to the low puddle density limit, $ r_p/R \ll 1$.  When density is increased, higher order terms in the multipole expansion become non-negligible. The resulting problem is clearly much more complex, with frustrated couplings that (at least in $d>2$) could lead to a gauge-glass phase for $r_p/R$ greater than a critical value, as conjectured in a related study of $d$-wave puddles~\cite{KivelsonSpivak}.

However, even in the small concentration limit, there are inevitably rare regions where the $(0,0)$ coupling is not the dominant coupling, and frustration can occur. To determine whether such frustrated regions significantly affect our conclusions, we must estimate their density: namely, the probability of finding nearby pairs of puddles where the $(0,0)$ coupling is the same order as the $(1,0)$ and $(0,1)$ couplings. A specific model of puddles which are almost spherical, apart from order-$Q^{-1}$ fluctuations of their boundaries, is analyzed in detail in the SM below. However the answer can be guessed on dimensional grounds. With $Q^{-1}$ fixed, we ask how the typical sizes of $|\eta^{0}|$ and $|\bm \eta^1|$ scale as the puddle size $r_p$ is varied. (The averages of these quantities vanish if translation and inversion symmetry are preserved on average in the disordered system.) The scaling of each is \textit{a priori} nontrivial, but inspecting Eq.~\eqref{eq:etaGen} tells us that the ratio of the widths of distributions must scale like $r_p$. Indeed, for the specific puddle model used in the SM below, $\eta^0 \sim r_p^{(d-1)/2} $ and $|\bm \eta^1| \sim r_p^{(d+1)/2}$.

Observe in Eq.~\eqref{eq:expansion} that the $(1,0)$ and $(0,1)$ Josephson couplings come with an extra factor $1/R$. The distribution of the $(0,1)$ contribution to a given $J_{ij}$ is therefore narrower than the $(0,0)$ one by a factor of $\sim r_p/R$, and we expect the two contributions to be of the same order with probability $\sim r_p/R$. This assumes that there are no strong positive correlations between $\eta^0$ and $\bm{\eta}^1$, which we show is true for the model in the SM below.

Thus, after fixing the gauge appropriately, we have a random-exchange XY model with a concentration $\sim r_p/R$ of near neighbor antiferromagnetic bonds, which are a factor of $\sim r_p/R$ smaller than the typical ferromagnetic coupling. Given that the frustrating effects are both rare and small in powers of $r_p/R$, the macroscopic $s$-wave phase will be stable, well away from the asymptotic $r_p/R \rightarrow 0$ limit. Furthermore, any collective glassy physics that might result from these frustrated regions lives at temperature scales that are many factors of $r_p/R$ down from the global superconducting transition. When $L_T \gg R$, the long-range nature of the interactions is expected to further diminish the  influence of the rare frustrated couplings.

\subhead{Acknowledgments}
We are grateful to Andrew C. Yuan, Matthew C. O'Brien, J\"org Schmalian, J. C. S\'eamus Davis, and Eduardo Fradkin for useful discussions. J. M. M. was supported by a Leinweber Institute for Theoretical Physics postdoctoral fellowship at Stanford University. Z. D. S. was supported by a Leinweber Institute for Theoretical Physics postdoctoral fellowship at Stanford University and in part by the Gordon and Betty Moore Foundation EPiQS initiative under Grant No.~GBMF8686.01. A. P. and S. A. K. were supported in part by NSF-BSF Award No.~DMR-2310312 at Stanford.

\subhead{Data availability}
There are no publicly available research data or software supporting this manuscript. Requests for further information or data should be sent to the authors.

\bibliography{references}

\clearpage
\onecolumngrid

\begin{center}
\textbf{\large Supplemental Material for ``Charge-$2e$ superconductivity from a disordered pair density wave"}
\end{center}

\section{Coupling of PDW puddles to macroscopic superconductors}\label{App:PDWtoUSC}
In this Appendix, we will consider the coupling between a PDW puddle and a macroscopic, clean, uniform $s$-wave superconducting lead. We take the lead to be semi-infinite, occupying all $x<0$, and with a cross sectional area that is macroscopically large.  Similar to before, the couplings between the puddles and the leads will be mediated by the propagation of Cooper pairs through the Fermi-liquid. The Fermi-liquid electrons couple to the boundary of the lead via
\begin{equation}
    H_{\text{lead-tun}} = g \int d^{d-1} \bm{r}_{\parallel} \Delta_{\text{lead}} \psi^\dagger_{\uparrow}(\bm{r}_{\parallel})\psi^\dagger_{\downarrow}(\bm{r}_{\parallel}) + h.c.,
\end{equation}
where $\psi^\dagger_{\sigma}$ are the electron operators, $\bm{r}_{\parallel} = (0, y,\dots)$ is the position coordinate along the boundary of the lead, $g$ is the tunneling amplitude between the lead and the Fermi-liquid, and $\Delta_{\text{lead}}$ is the superconducting order parameter of the lead.

The Josephson coupling between the lead and the $i^\text{th}$ PDW puddle is again found by integrating out the Fermi-liquid perturbatively in both PDW amplitude $|\Delta_i|$, and the couplings to the lead $g$. Provided that the electrons propagate as $1/r^d$ in the semi-infinite geometry (for $r \ll L_T$), the Josephson coupling is
\begin{equation}
    J_{\text{lead},i} = 4\nu g |\Delta_{\text{lead}}| |\Delta_i| \int d^{d-1}\bm{r}_\parallel \int d^d \bm{r}_i \frac{f(\bm{r}_i)\cos(\bm{Q}_{i}\cdot \bm{r}_i + \alpha_i/2)}{|\bm{R}_i + \bm{r}_i - \bm{r}_\parallel|^d},
\label{app:JC_PDWMacro}\end{equation}
provided that $|\bm{R}_{i,\perp}|\ll L_T$, where $\bm{R}_{i,\perp}$ is the shortest vector connecting $\bm{R}_i$ to the edge of the lead. When $|\bm{R}_{i,\perp}|\gtrsim L_T$ the Josephson coupling is exponential suppressed and can be ignored.

First, let us consider the couplings to puddles that are far away, $|\bm{R}_{i,\perp}| \gg r_p $. For such puddles, we can expand the Josephson coupling in terms of the multipoles of the PDW puddle. In $d =3$, the leading order terms in this expansion are
\begin{equation}
    J_{\text{lead},i} = 4 \nu g |\Delta_{\text{lead}}||\Delta_i| \left[ \frac{2\pi \eta^0_{i}} {|\bm{R}_{i,\perp}|} - \frac{2\pi {\bm \eta}^1_{i}\cdot \bm{n}_{i,\perp}} {|\bm{R}_{i,\perp}|^2} + \dots \right],
\end{equation}
and in $d =2$, they are
\begin{equation}
    J_{\text{lead},i} = 4 \nu g |\Delta_{\text{lead}}||\Delta_i| \left[ \frac{\pi \eta^0_{i}} {|\bm{R}_{i,\perp}|} - \frac{\pi {\bm \eta}^1_{i}\cdot \bm{n}_{i,\perp}} {|\bm{R}_{i,\perp}|^2} + \dots \right],
\end{equation}
where $\eta^0$ and $\bm{\eta}^1$ are defined as in the main text, and $\bm{n}_{i,\perp} = \bm{R}_{i,\perp}/|\bm{R}_{i,\perp}|$. Here, we have assumed that the cross-sectional area of the lead $A_{\text{lead}}$ is much larger than $|\bm{R}_{i,\perp}|^{d-1}$.

For puddles that are close to the macroscopic $s$-wave superconductor, $|\bm{R}_{i,\perp}| \sim r_p$, one must consider all terms in the multipole expansion. In general, the signs of higher order terms will be uncorrelated with the signs of the $\eta^0$ term. This can be understood by considering how the different terms transform when $\bm{r}_i$ and $\bm{Q}_i$ are rotated. $\eta^0_i$ is invariant under any such transformations, but the higher order terms generically transform non-trivially. The signs of the couplings to the close-by puddles will therefore be random and uncorrelated with the signs of the distant puddles.

\section{Estimates for nearly spherical puddles}\label{app:nearly_spherical}

In this Appendix, we will consider the joint distribution of $\eta^0$ and $\bm \eta^1$ for puddles that are nearly spherical, but with order $Q^{-1}$ fluctuations of the boundary:
\begin{equation}
    f_i(\mathbf{r}) = \begin{cases} 1 & \text{ for } |\mathbf{r}| < r_{p,i} + \delta\rho_i(\Omega),\\
    0 & \text{ otherwise},
    \end{cases}
\label{eq:puddleShapeDef}\end{equation}
where $r_{p,i}$ is the order-$r_p$ radius of the spherical reference of puddle $i$, and $\delta\rho_i(\Omega)$ encodes the order-$Q^{-1}$ radial fluctuations around this spherical reference.  The fluctuations of $\delta \rho_i$ are symmetric about $0$, with correlation length along the puddle of order $Q^{-1}$. In this section, we will assume that the puddles contain many PDW periods, $Q^{-1} \ll r_p$.

\subsection{Spherical puddles}

Using Eq.~\eqref{eq:puddleShapeDef}, we can divide up the contributions to $\eta^0$ and $\bm{\eta}^1$ into two pieces: those arising from the integration over the spherical reference, and those arising from the boundary fluctuations.
Let us first calculate the spherical contribution.

For a spherical puddle in $d = 3$, $\eta^0$ is
\begin{equation}\begin{split}
\eta_{i}^0&=\int_{r \leq r_{p,i}} d^3\bm{r} \cos(\bm{Q}_{i}\cdot \bm{r} + \alpha_i/2)\\
&= 4\pi \cos(\alpha_i/2) \frac{(\sin(|\bm{Q}_{i}|r_{p,i}) - |\bm{Q}_{i}|r_{p,i} \cos(|\bm{Q}_{i}|r_{p,i}))}{|\bm{Q}_{i}|^3} \rightarrow  -r_{p,i} 4\pi \cos(\alpha_i/2)\frac{\cos(|\bm{Q}_{i}|r_{p,i})}{|\bm{Q}_{i}|^2},
\end{split}\end{equation}
and $\bm{\eta}^1$ is
\begin{equation}\begin{split}
\bm{\eta}^1_i &=\int_{r\leq r_{p,i}} d^3\bm{r} \bm{r} \cos(\bm{Q}_{i}\cdot \bm{r} + \alpha_i/2)\\
&= 4\pi \sin( \alpha_i/2) \frac{3 |\bm{Q}_{i}|r_{p,i} \cos(|\bm{Q}_{i}|r_p) + (|\bm{Q}_{i}|^2r_{p,i}^2 - 3) \sin(|\bm{Q}_{i}|r_{p,i}) }{|\bm{Q}_{i}|^4}\bm{n}_{\bm{Q}_i}\rightarrow r_{p,i}^24 \pi \sin( \alpha_i/2) \frac{  \sin(|\bm{Q}_{i}|r_{p,i}) }{|\bm{Q}_{i}|^2}\bm{n}_{\bm{Q}_i}
\end{split}\end{equation}
where $\bm{n}_{\bm{Q}_i} = \bm{Q}_i/|\bm{Q}_i|$ and ``$\rightarrow$" reflects the $r_{p,i}\gg Q^{-1}$ limit. Note that when $\cos(\alpha_i/2) = 0$, $\eta^0$ vanishes, while $\bm{\eta}^1 = \pm r_{p,i}^24 \pi\frac{  \sin(|\bm{Q}_{i}|r_{p,i}) }{|\bm{Q}_{i}|^2}\bm{n}_{\bm{Q}_i}$, which is generically non-zero.

For spherical puddles in $d = 2$, $\eta^0$ is
\begin{equation}\begin{split}
\eta^0_i = \frac{2\pi r_{p,i} \cos(\alpha_i/2)J_1(|\bm{Q}_{i}|r_{p,i})}{|\bm{Q}_{i}|} \rightarrow  (r_{p,i})^{1/2}\frac{\sqrt{8\pi}\cos(\alpha_i/2)\sin(|\bm{Q}_{i}|r_{p,i} - \pi/4)}{|\bm{Q}_{i}|^{3/2} }
\end{split}\end{equation}
and $\bm{\eta}^1$ is
\begin{equation}\begin{split}
\bm{\eta}^1_i = - \frac{2\pi r^2_{p,i}  \sin(\alpha_{i}/2) J_2(|\bm{Q}_{i}|r_{p,i}) }{|\bm{Q}_{i}|}\bm{n}_{\bm{Q}_i} \rightarrow (r_{p,i})^{3/2} \frac{\sqrt{8\pi}\sin(\alpha_i/2)\sin(|\bm{Q}_{i}|r_{p,i} + \pi/4)}{(|\bm{Q}_{i}|)^{3/2}}\bm{n}_{\bm{Q}_i}
\end{split}\end{equation}
where $J_n$ are the Bessel functions. Again, when $\cos(\alpha_i/2) = 0$, $\eta^0$ vanishes, while $\bm{\eta}^1$ is non-zero (up to fine tuning).

To summarize: these contributions are random, due to the random field $\alpha_i$ on each puddle and the randomness in the radius of the puddle $r_{p,i}$. For $r_{p,i} \gg Q^{-1}$, the contributions to $\eta^0$ from the spherical part of the puddle scales as $r_p$ in $d = 3$ and as  $r^{1/2}_p$ in $d = 2$.  In the same limit, $\bm{\eta}^1$ scales as $r^2_p$ in $d = 3$ and as  $r^{3/2}_p$ in $d = 2$, as one would expect from dimensional analysis. The spherical contributions to $\eta^{0}$ and $\bm{\eta}^{1}$ are highly correlated random variables: for example, their dependence on $\alpha_i$ is $\cos(\alpha_i/2)$ and $\sin(\alpha_i/2)$ respectively. However, the correlation between their magnitudes is \textit{negative}: when $|\eta^0|$ is unusually small $|\bm{\eta}^1|$ is of the same order as its typical value. Thus, in what follows we may as well consider the spherical contributions to $\eta^{0}$ and $\bm{\eta}^{1}$ as independent.

\subsection{Boundary fluctuations}

Let us now consider the contribution from radius fluctuations, which is uncorrelated with the spherical part considered above. This contribution is independent of the details of how we model the PDW puddles, in the sense that the following does not depend on the choice of having almost spherical puddles. The following analysis applies to any puddles where the puddle boundaries are asymptotically smooth. Consider the relevant contribution to the integral that defines $\eta^0$.
If we perform the integral over the radial coordinate, we are left with a $(d-1)$-dimensional surface integral over an order-$Q^{-1}$ function. This function will be symmetric around zero, and spatially uncorrelated over angles of order $Q^{-1}/r_{p}$, as the radius fluctuations are uncorrelated with the cosine term in $\eta^0$. Based on this, the surface integral will have typical value of order $r_p^{(d-1)/2}$. A similar argument indicates that the contribution to $\bm{\eta}^1$ from the fluctuating boundary is of order $r_p^{(d+1)/2}$.

To be more concrete, let us consider a region of the puddle boundary, $\mathcal{B}$ of size $\sim r^{d-1}_p$ that is parallel to $\mathbf{Q}$ (we expect that such regions dominate the radius-fluctuation contribution to $\eta^0$). Consider an auxiliary model of a ``height'' field $h(\mathbf{x})$ which is a proxy for radius fluctuations, with $\mathbf{x}$ representing a tangential coordinate on the sphere. Measuring distances in units of $Q^{-1}$, the height field satisfies
\begin{equation}
    \overline{h(\mathbf{x})} = 0, \quad \overline{h(\mathbf{x})h(\mathbf{x}')} =  e^{-\rm{cst.}|\mathbf{x}-\mathbf{x'}|}.
\end{equation}
In terms of $h$, the relevant contribution to $\eta_0$ can be approximated as
 \begin{equation}
     \eta_0 \sim \int_{\mathcal{B}} d^{d-1} x\,  h(\mathbf{x}) \cos x_1.
 \end{equation}
This is given as a sum over order-$r_p^{d-1}$ uncorrelated symmetrically distributed random variables. This contribution vanishes on average, $\overline{\eta^0}=0$, but has typical value
 \begin{equation}
     \sqrt{\overline{(\eta^0)^2}} \sim r_p^{(d-1)/2}.
 \end{equation}
We can do a similar estimate for $\bm \eta^1$, by weighting the random sum with coordinates. Again, $\overline{\bm \eta^1} = 0$, but now the typical value is
  \begin{equation}
      \sqrt{\overline{(\bm \eta^1)^2}} \sim r_p^{(d+1)/2},
  \end{equation}
  up a factor of $r_p$ from that of $\eta^0$, as could have been expected by dimensional analysis. Furthermore, for this simple model via the height field, $\eta^0$ and $\bm{\eta}^1$ are independent (essentially on symmetry grounds). In particular, their covariance vanishes.

\section{Connection with $\mathbb{Z}_2$ lattice gauge theory}\label{app:Z2gauge}

In this Appendix, we elaborate on the relationship between the microscopic model of PDW puddles studied in the main text and an effective $\mathbb{Z}_2$ lattice gauge theory (coupled to matter fields). We will begin by reviewing the general formulation of a disordered $\mathbb{Z}_2$ gauge theory and discuss its phase diagram. Then we will derive the specific $\mathbb{Z}_2$ gauge theory that arises from a low energy approximation of the microscopic model with PDW puddles.

\subsection{General $\mathbb{Z}_2$ gauge theory phase diagram with two scalar matter fields}\label{subsec:Z2_abstract}

We first give a brief review of $\mathbb{Z}_2$ gauge theory coupled to a pair of $2\pi$-periodic compact scalar fields $\theta, \phi$ in the presence of disorder. With applications to PDW superconductors in mind, we should think about $\theta$ as the phase of $\Delta_0$ and $\phi$ as the phase of $\rho_{\bm{Q}}$ such that the PDW order parameters are written as $\Delta_{\pm \bs{q}} = \Delta_0 \rho_{\pm \bs{Q}} = e^{i \theta \pm \phi}$. In this decomposition of $\Delta_{\pm \bs{Q}}$, $\theta$ and $\phi$ both carry gauge charge under the $\mathbb{Z}_2$ gauge field. Moreover, we take $\rho_{\bs{Q}}$ to be neutral and $\Delta_0$ to have charge 2 under the external electromagnetic gauge field $A$. The most general low energy effective Hamiltonian in the absence of disorder then takes the form
\begin{equation}\label{eq:abstract_Z2}
    H = H[\theta, 2A - a] + H[\phi, a] + H[a] \,,
\end{equation}
where $H[a]$ is a kinetic term for the dynamical $\mathbb{Z}_2$ gauge field $a$ (e.g. products of $U_{ij} \equiv e^{ia_{ij}}$ operators around lattice plaquettes) and
\begin{equation}
    H[\theta, 2A - a] = \sum_{ij} J_{\theta} \cos(\theta_i - \theta_j - 2A_{ij} + a_{ij}) \,, \quad H[\phi, a] = J^-_{\phi} \sum_{ij} \cos(\phi_i - \phi_j - a_{ij}) \,.
\end{equation}
On top of this clean Hamiltonian, there are several types of disorder that one can add. First, we must include a random field disorder that pins the $2\bs{Q}$ CDW order. This type of disorder enters the Hamiltonian as
\begin{equation}
    H_{\rm random-field} = \sum_i h_i \cos (2 \phi_i - \alpha_i) \,,
\end{equation}
where $h_i$ and $\alpha_i \in [-\pi, \pi)$ are quenched random variables. Furthermore, since translation symmetry is explicitly broken by the disorder potential, we should no longer demand that the Hamiltonian be invariant under $\phi_i \rightarrow \phi_i + \alpha$. This means that we have to add an additional (random in sign) inter-puddle cosine term $\sum_{ij} \cos (\phi_i + \phi_j + a_{ij})$ on top of the standard cosine term in a gauged XY model $\sum_{ij} \cos (\phi_i - \phi_j + a_{ij})$. Finally, within each cosine term, we can add a random force that locally pins $\phi_i - \phi_j$ or $\phi_i + \phi_j$ to a random orientation. Putting all of these terms together, we obtain the general disordered Hamiltonian
\begin{equation}\label{eq:abstract_Z2_disordered}
    \begin{aligned}
    H_{\rm dis} &= H[a] + \sum_{ij} J_{\theta, ij} \cos(\theta_i - \theta_j - 2 A_{ij} + a_{ij}) + \sum_i h_i \cos(2\phi_i - \alpha_i) \\
    &+ J_{\phi,ij}^- \sum_{ij} \cos(\phi_i - \phi_j + a_{ij} + \eta^-_{ij}) + J_{\phi,ij}^+ \sum_{ij} \cos(\phi_i + \phi_j + a_{ij} + \eta^+_{ij})  \,.
    \end{aligned}
\end{equation}
Note that we can always define our couplings so that $J_{\theta, ij}, J^{\pm}_{\phi, ij}$ have random amplitudes but fixed sign.

From here, we can analyze the phase structure of $H_{\rm dis}$. We will focus on the limit of strong pinning $h_i \gg 1$ as it is most relevant to the model studied in the main text. We will briefly comment on the weak pinning limit towards the end.

\begin{center}
    Strong pinning limit $h_i \gg 1$
\end{center}
In the strong pinning limit, the continuous variable $\phi$ reduces to an Ising variable up to an onsite random phase rotation $\phi_i \rightarrow \phi_i + \alpha_i/2$. If we represent the Ising variable by $\tau_i$, then the full theory becomes
\begin{equation}
    H_{\rm dis} = H[\theta, 2A - a] + H[\tau, a] + H[a] \,,
\end{equation}
where $H[\theta, 2A - a], H[a]$ are unmodified and $H[\tau, a]$ takes the form
\begin{equation}
    \begin{aligned}
    H[\tau, a] = \sum_{ij} J_{\tau, ij} \tau_i \tau_j e^{ia_{ij}} \,, \quad J_{\tau,ij} = J_{\phi,ij}^- \cos(\frac{\alpha_i - \alpha_j}{2} + \eta^-_{ij}) + J_{\phi,ij}^+ \cos(\frac{\alpha_i + \alpha_j}{2} + \eta^+_{ij}) \,.
    \end{aligned}
\end{equation}
The fate of this theory depends crucially on the dynamics of $\tau$. If the random field couplings and random force couplings are sufficiently strong so that the effective coupling $J_{\tau, ij}$ contains a sufficiently high density of frustrated bonds, then $\tau$ goes into a spin glass phase. In the limit of maximal frustration, where $J_{\tau, ij}$ becomes uncorrelated random variables that are symmetrically distributed around 0, the $\tau$ sector reduces to the standard Edwards-Anderson Ising spin glass model, as considered in Ref.~\cite{MrossSenthil}. However, we wish to be more general in this section and consider different possible fates of the $\tau$ sector.

If $\tau$ is in a disordered phase (either a trivial disordered phase or an Ising spin glass), then there are generally two possible fates for the full Hamiltonian: (1) If $J_{\theta} \ll 1$, then $\theta$ does not order. The resulting electronic phase does not break any symmetry but may have glassy order. In $d = 3$, if the gauge field kinetic coupling $K$ is not strongly disordered, it is also possible for the model to realize a classical topological order, which is the classical version of the I* phase in Senthil and Fisher's theory of quantum $\mathbb{Z}_2$ fractionalization in 2+1 dimensions~\cite{Senthil2000_Z2frac}. (2) If $J \gg 1$, then $\theta$ enters an ordered phase. Since $\theta_i$ is not gauge-invariant, the minimally gauge-invariant order parameter is $2 \theta_i$ and we obtain a charge $4e$ superconductor.

The physics is drastically different if $\tau$ instead enters an ordered phase. The ordering of $\tau$ Higgses the $\mathbb{Z}_2$ gauge field $a_{ij}$. Once Higgsing occurs, $H[\theta, 2A]$ simply describes an XY model for the composite field $e^{i\theta_i} \tau_i$, which carries charge 2e. When the XY model is disordered, we have either a trivial phase or a glassy phase for $\theta$; when it is ordered, we have a charge $2e$ superconductor rather than a charge $4e$ superconductor. Note that this charge $2e$ superconductor only has a single gapless Goldstone mode, in contrast to the clean PDW which has two distinct Goldstone modes corresponding to $\Delta_{\bs{Q}}$ and $\Delta_{-\bs{Q}}$. As a result, the ground state of the charge $2e$ superconductor is indistinguishable from a disordered charge $2e$ s-wave superconductor.

\begin{center}
    Weak pinning limit $h_i \ll 1$
\end{center}

The weak pinning limit is more intricate. For an ordinary XY model, a weak pinning term is a relevant perturbation that grows under the renormalization group flow. If this flow continues all the way to strong pinning, then we can simply apply the arguments in the previous section. However, in the presence of random force couplings $\eta^{\pm}_{ij}$, the renormalization group flow of $h_i$ may be strongly affected and a more careful analysis is needed to determine the correct low energy fixed point. Depending on the ratio between microscopic couplings, the model may flow to the strong pinning limit in which our previous analysis applies, or it may flow to a disordered phase (possibly with spin glass order). Since the weak pinning limit is not relevant to our model of PDW puddles, we will not discuss this renormalization group flow in more detail and refer interested readers to the detailed analysis in Ref.~\cite{MrossSenthil} (which includes $\eta^-_{ij}$ but not $\eta^+_{ij}$).

\subsection{Deriving a $\mathbb{Z}_2$ gauge theory from the puddle Hamiltonian}\label{subsec:Z2_from_puddle}

Having reviewed the phase diagram of the disordered $\mathbb{Z}_2$ gauge theory, let us see how it arises as a low energy approximation of the microscopic model of PDW puddles studied in the main text.

The microscopic Hamiltonian for disordered PDW puddles starts out with only three terms:
\begin{equation}
    \begin{aligned}
    H &= \sum_i h_i e^{i\alpha_i} \Delta^*_{i, \bm{Q}} \Delta_{i, - \bm{Q}} - \sum_{ij} J_{ij} \Delta_{i,\bm{Q}}^* \Delta_{j,\bm{Q}} + \sum_{ij} J_{ij} \Delta_{i,\bm{Q}}^* \Delta_{j,-\bm{Q}} + h.c. \\
    &= \sum_i h_i e^{i\alpha_i} \Delta^*_{i, \bm{Q}} \Delta_{i, - \bm{Q}} - \sum_{ij} J_{ij} \Delta_{i,0}^* \Delta_{j,0} \left[\rho_{i, \bm{Q}}^* \rho_{j, \bm{Q}} +  \rho_{i, \bm{Q}}^*  \rho_{j, -\bm{Q}}\right] + h.c. \,.
    \end{aligned}
\end{equation}
Since the last line arises from rewriting $\Delta_{i,\bm{Q}}$ in terms of redundant variables $\Delta_{i, \bm{Q}} = \Delta_{i,0} \rho_{i, \bm{Q}}$, there is an exact local $\mathbb{Z}_2$ redundancy $\Delta_{i,0} \rightarrow - \Delta_{i,0}$ and $\rho_{i,\bm{Q}} \rightarrow - \rho_{i, \bm{Q}}$, although there is no apparent dynamical $\mathbb{Z}_2$ gauge field.

Now we introduce the following Hubbard-Stratanovich transformation:
\begin{equation}
    - \sum_{ij} J_{ij} \Delta^*_{i,0} \Delta_{j,0} \rho^*_{i,\bm{Q}} \rho_{j,\bm{Q}} \rightarrow \sum_{ij} \left[\frac{1}{2} J_{ij} |\chi_{ij}|^2 + J_{ij} (\chi_{ij} \Delta^*_{i,0} \Delta_{j,0} + \chi^*_{ij} \rho^*_{i,\bm{Q}} \rho_{j,\bm{Q}})\right] \,,
\end{equation}
\begin{equation}
    - \sum_{ij} J_{ij} \Delta^*_{i,0} \Delta_{j,0} \rho^*_{i,\bm{Q}} \rho_{j,-\bm{Q}} \rightarrow \sum_{ij} \left[\frac{1}{2} J_{ij} |\tilde \chi_{ij}|^2 + J_{ij} (\tilde \chi_{ij} \Delta^*_{i,0} \Delta_{j,0} + \tilde \chi^*_{ij} \rho^*_{i,\bm{Q}} \rho_{j,-\bm{Q}})\right] \,.
\end{equation}
Solving the saddle point equations for $\chi_{ij}$ and $\tilde \chi_{ij}$ gives a uniform mean-field in which $\ev{\chi_{ij}} = \ev{\tilde \chi_{ij}} \in \mathbb{R}$. As usual, we assume that amplitude fluctuations of $\chi_{ij}, \tilde \chi_{ij}$ are unimportant and focus on the sign fluctuations parametrized by $\chi_{ij} = \tilde \chi_{ij} = \chi_0 U_{ij}$, where $U_{ij} \in \pm 1$. Further neglecting gapped amplitude fluctuations of the PDW order parameter, we can parametrize $\Delta_{i,0}$ and $\rho_{i,\pm \bm{Q}}$ by phase variables $\Delta_{i,0} \sim e^{i\theta_i}$ and $\rho_{i, \bm{Q}} \sim e^{i\phi_i}$. The final result is a Hamiltonian of the form
\begin{equation}
    H_{\rm eff} = \sum_i h_i \cos(2 \phi_i - \alpha_i) + \sum_{ij} \left[2J_{ij} \chi_0 U_{ij} e^{i(\theta_i - \theta_j)} + J_{ij} \chi_0 e^{-i(\phi_i - \phi_j)} + J_{ij} \chi_0 e^{-i(\phi_i + \phi_j)}\right]
\end{equation}
Using the Hermiticity of the coupling matrix $J_{ij}$, we can simplify the Hamiltonian to the more familiar form
\begin{equation}\label{eq:derived_Z2}
    H_{\rm eff} = \sum_i h_i \cos(2 \phi_i - \alpha_i) + \sum_{ij} J_{ij} \chi_0 U_{ij} \left[2\cos (\theta_i - \theta_j) + \cos(\phi_i - \phi_j) + \cos(\phi_i + \phi_j) \right] \,,
\end{equation}
where $U_{ij}$ is recognized as a dynamical $\mathbb{Z}_2$ gauge field.

From \eqref{eq:derived_Z2}, we see that the low energy limit of the model of PDW puddles reduces to the abstract model considered in \eqref{eq:abstract_Z2_disordered} with an appropriate choice of couplings. First, note that the Hamiltonian for the $\phi$ sector in \eqref{eq:derived_Z2} matches the description in \eqref{eq:abstract_Z2_disordered} if we set $J^{\pm}_{\phi} = J_{ij} \chi_0$, $J_{\theta} = 2J_{ij} \chi_0$, $U_{ij} = e^{ia_{ij}}$ and $\eta^{\pm}_{ij} = 0$. The equality between $J^-{\phi}$ and $J^+_{\phi}$ is an important feature of the PDW puddle model which reduces the amount of frustration in the effective Hamiltonian for $\phi$. Furthermore, note the dynamical $\mathbb{Z}_2$ gauge field does not have a kinetic term $H[a]$ in \eqref{eq:derived_Z2}. To generate an effective kinetic term, we can coarse-grain the system over larger and larger length scales and run the renormalization group flow. Truncating the flow at some finite energy cutoff $\Lambda$ generally leads to an effective model in which all the effective couplings are $\mathcal{O}(1)$ constants (including the gauge kinetic coupling). In this way, we can recover the model in \eqref{eq:abstract_Z2_disordered} with an appropriate UV cutoff set by $\Lambda$.

In the strong pinning limit $h_i \gg J_{ij}$, $e^{i\phi_i}$ reduces to an effective Ising variable $\tau_i e^{i\alpha_i/2}$. The Hamiltonian then simplifies to
\begin{equation}
    H_{\rm eff} = \sum_{ij} 2J_{ij} \chi_0 U_{ij} \cos (\theta_i - \theta_j) + \sum_{ij} 2 J_{ij} U_{ij} \tau_i \tau_j \cos(\alpha_i/2) \cos(\alpha_j/2) \,.
\end{equation}
Without loss of generality, we can choose $\alpha_i, \alpha_j$ to take value in $[-\pi,\pi)$ such that $\cos(\alpha_i/2) > 0$ for all possible $\alpha_i$. In that case, minimizing the second term sets $U_{ij} \tau_i \tau_j = 1$ for all $i,j$. This effectively Higgses $U_{ij}$ and further simplifies the model to
\begin{equation}
    H_{\rm eff}[\{\theta_i, \tau_i\}] = \sum_{ij} 2 J_{ij} \chi_0 \tau_i \tau_j \cos (\theta_i - \theta_j) \,,
\end{equation}
which is precisely the effective model in the main text, derived from accounting for inter-puddle Cooper pair hopping mediated by the intervening Fermi liquid. The formation of a charge $2e$ superconductor in this scenario agrees with the general analysis of disordered $\mathbb{Z}_2$ gauge theories presented in App.~\ref{subsec:Z2_abstract}.

As a final comment, if we perturb the microscopic couplings so that $J^+_{\phi} \neq J^-_{\phi}$ but $|J^+_{\phi} - J^-_{\phi}| \ll |J^+_{\phi}|, |J^-_{\phi}|$, a small amount of frustration is introduced into the $\tau$ sector. However, since an Ising ferromagnet is stable to weak frustrated random couplings, $\tau$ continues to order and the charge $2e$ superconductivity persists. Thus, the phase of matter obtained in the main text is stable.

\end{document}